\documentclass[twocolumn,showpacs,preprintnumbers,showkeys,superscriptaddress]{revtex4}
\usepackage{graphicx}
\usepackage{dcolumn}
\usepackage{bm}

\def\eqref#1{Eq.~(\ref{eq:#1})}
\def\eqlab#1{\label{eq:#1}}
\def\figref#1{Fig.~(\ref{fig:#1})}
\def\figlab#1{\label{fig:#1}}
\def\tabref#1{Table \ref{tab:#1}}
\def\tablab#1{\label{tab:#1}}
\def\secref#1{Section~\ref{sec:#1}}
\def\seclab#1{\label{sec:#1}}

\begin{document}

\title{Patterns of the ground states
in the presence of random interactions: nucleon systems}
\author{Y. M. Zhao}     \email{ymzhao@riken.jp}
\affiliation{Cyclotron Center,  Institute of Physical Chemical Research (RIKEN), \\
Hirosawa 2-1, Wako-shi,  Saitama 351-0198,  Japan}
\affiliation{Department of physics,  Southeast University, Nanjing 210018 China}
\author{A. Arima}
\affiliation{The House of Councilors, 2-1-1 Nagatacho,
Chiyodaku, Tokyo 100-8962, Japan}
\author{N. Shimizu}     \email{shimizu@nt.phys.s.u-toyo.ac.jp}
\affiliation{Department of Physics, University of Tokyo, Bunkyo-ku,
Tokyo 113-0033, Japan}
\author{K. Ogawa}     \email{ogawa@physics.s.chiba-u.ac.jp}
\affiliation{Department of physics, Chiba university,
Yayoi-cho 1-33, Inage, Chiba 263-8522, Japan}
\author{N.~Yoshinaga}     \email{yosinaga@phy.saitama-u.ac.jp}
\affiliation{Department of Physics, Saitama University,   Saitama 338 Japan}
\author{O. Scholten}     \email{scholten@kvi.nl}
\affiliation{Kernfysisch Versneller Instituut, University of Groningen,
9747 AA Groningen, The~Netherlands}
\date{today}

\begin{abstract}

We present our results on properties of ground states for nucleonic systems in 
the presence of random two-body interactions. In particular we present
probability distributions for parity, seniority, spectroscopic
(i.e.,  in the laboratory framework) quadrupole
moments and $\alpha$ clustering in the ground states. We find that the
probability distribution for the parity of the ground states obtained by a
two-body random ensemble simulates that of realistic nuclei
with $A\ge 70$: positive parity is 
dominant in the ground states of even-even nuclei while for odd-odd nuclei and
odd-mass nuclei we obtain with almost equal probability ground states with
positive and negative parity. In addition we find that for the ground states,
assuming pure random interactions, low seniority is not favored,  no dominance
of positive values of spectroscopic quadrupole deformation is
observed, and there is no sign of 
$\alpha$-clustering correlation, all in sharp contrast to realistic nuclei. 
Considering a mixture of a random and a realistic interaction, we observe a
second order phase transition for the $\alpha$-clustering correlation probability.

\end{abstract}

\maketitle

\section{\seclab{Intro}Introduction}

It was discovered in Ref. \cite{Johnson1} that the dominance 
of  spin zero ground states (0 g.s. ) can be obtained by diagonalizing 
a scalar two-body  Hamiltonian with random valued matrix elements, 
a so-called two-body random ensemble (TBRE) Hamiltonian.  The 
0 g.s. dominance was soon confirmed in Ref. \cite{Bijker0} for sd-boson systems. These 
feature was found to be robust and insensitive to the detailed statistical properties 
of the random Hamiltonian, suggesting that the 0 g.s. dominance 
arises from a very large ensemble of two-body interactions other than 
a simple monopole paring interaction.   An understanding of this discovery 
is very important, because this observation seems to be contrary to what is 
traditionally assumed in nuclear physics, where the 0 g.s. dominance 
in even-even nuclei is usually explained as a reflection of attractive pairing 
interaction between like nucleons.

There are a lot of efforts to understand this observation  but a fundamental 
understanding  is still out of reach \cite{Bijker}.
There are also many works \cite{Zhao0} studying other 
robust phenomena of many-body systems in the presence of the TBRE.  For 
example,  the studies of odd-even staggering of binding energies, generic 
collectivity, the behavior of energy centroid of fixed spin states, 
correlation, etc. 

The purpose of the present paper is to focus  our 
attention on some physical quantities for the ground states 
which have not been studied, specifically, parity, seniority, 
spectroscopic quadrupole moments (i.e., measured in the laboratory framework),
and $\alpha$-clustering  probability. For realistic nuclei these quantities show a
very regular pattern. In this paper we shall discuss
 whether  these regular patterns
are   robust in the presence of  random interactions.

As well known, all  even-even nuclei have
positive parity  ground states (i.e., $100\%$) 
whereas the ground states of nuclei with odd mass numbers  
 have only a slightly higher  
probability for positive parity than for negative parity. Odd-odd nuclei 
have almost equal probabilities for positive and negative-parity ground  
states($\sim 50\%$). The statistics for the  ground state parity of  
nuclei with mass number $A\ge 70$ are summarized  in \tabref{1}. As the first 
subject we will study the ground-state parity distribution using random
interactions.

\begin{table}
\caption{\tablab{1}
 The positive parity distribution of the ground states
of atomic nuclei.  We included all ground state parities of nuclei with mass
number $A \ge 70$. The data are taken from Ref.~\cite{Firestone}. We
haven't  taken into account  those nuclei for which the ground state parity
was not measured. }
\begin{ruledtabular}
\begin{tabular}{cccc}
counts & even-even & odd-$A$  & odd-odd  \\  \hline
verified (+)   & 487   & 281  & 118  \\
verified ($-$)   & 0     & 215  & 104  \\  \hline
tentative (+) & 0     & 159  &  70 \\
tentative ($-$) & 0     &  126 &  60
\end{tabular}
\end{ruledtabular}
\end{table}

The next subject that we shall discuss in this paper is  the distribution of
seniority in the ground states. Seniority~\cite{Racah} has been proven a very 
relevant concept in nuclear physics, in particular for spherical or
transitional nuclei. Seniority ($v$) is uniquely defined for a single-$j$
shell; it was generalized to the case of many-$j$ shells by Talmi in
Ref.~\cite{Talmi}. In Refs.~\cite{Johnson1,Johnson2} it  was reported that  the
pairing phenomenon seems favored simply as a consequence of the two-body nature
of the interaction. The ``pairing" of Refs.~\cite{Johnson1,Johnson2}  was
defined as a large matrix element of the $S$ pair annihilation operator between
the ground states of a $n$ fermion system to a $n-2$, $n-4$ $\cdots$ systems,
where the $S$ pair structure  is determined using the procedure of Talmi's
generalized seniority scheme. This indicated that the $S$-pair correlation is
dominant for the spin-0 g.s.\ of these systems. An examination of this
``pairing" correlation of fermions in a single-$j$ shell in Ref.~\cite{Zhao-3}
showed that, however, an enhanced probability for low seniority in the spin--0
g.s.\ is not observed in most of the calculations using a TBRE hamiltonian. For
many-$j$ shells, there have been few discussions to clarify this point so far.

Another subject that we shall discuss is the $\alpha$-clustering  correlation in
the presence of random interactions. The importance of the $\alpha$-clustering 
correlation  in light and medium nuclei have been emphasized by many
authors~\cite{Arima1}. The $\alpha$-clustering  correlation  also plays an important
role in astrophysical processes such as the Salpeter process in the formation
of $^{12}$C. Many  calculations of the low-lying states, using the
anti-symmetrized molecular dynamics model, have been done in recent
years~\cite{Enyo} to study the $\alpha$-clustering
and other clustering  correlations
for both stable and unstable light nuclei. $\alpha$-cluster  condensation was
suggested by Horiuchi, Schuck and collaborators in Ref.~\cite{Horiuchi}. As a
function of the mixing of a realistic and the TBRE interaction a phase
transition is observed in the $\alpha$-clustering  probability in the ground state.

In this paper we also  discuss   the spectroscopic quadrupole moments
(i.e., measured in the laboratory framework)
 of the ground states. A
positive value for spectroscopic
quadrupole  deformation is   dominant in the low-lying
states of atomic nuclei. Recently it has been argued in Ref.~\cite{Tajima} that
this is due to the interference of spin-orbit and $l^2$ terms of the Nilsson
potential.

Our calculations are based on the use of TBRE interactions. The single-particle
matrix elements are set to be zero. The Hamiltonian that we use conserves the
total angular momentum and isospin:
\begin{eqnarray}
&& H = \sum_{j_1 j_2 j_3 j_4}^{JT}
\sqrt{2J+1} \sqrt{2T+1} G_{j_1 j_2 j_3 j_4}^{JT}  \nonumber
\\
&&
\frac{1}{ \sqrt{1+ \delta_{j_1 j_2}} \sqrt{1+ \delta_{j_3 j_4}} } \nonumber
\\
&& \left(
\left(  a^{\dagger}_{j_1 t} \times a^{\dagger}_{j_2 t} \right)^{(JT)}
\times
\left(  \tilde{a}_{j_3 t} \times \tilde{a}_{j_4 t} \right)^{(JT)}
\right)^{(00)},
\end{eqnarray}
where these $G_{j_1 j_2 j_3 j_4}^{JT}$ are defined by $\langle j_1 j_2 JT |V|
j_3 j_4 JT \rangle$ and follow the following distribution,
\begin{equation}
\rho (G_{j_1 j_2 j_3 j_4}^{JT}) =    \frac{1}{\sqrt{2\pi x} }
{\rm exp}
\left( - \frac{ \left( G_{j_1 j_2 j_3 j_4}^{JT} \right)^2}{2x} \right)
\end{equation}
with
\begin{equation}
 x =
 \left\{ \begin{array}{ll}
 1              & {\rm if ~} |(j_1 j_2)JT \rangle = |(j_3j_4)JT \rangle \\
 \frac{1}{2}    & {\rm otherwise}
\end{array} \right. ~ . ~
\end{equation}
The Hamiltonian such defined is called a TBRE Hamiltonian. Here $j_1, j_2, j_3$
and $j_4$ denote the respective single-particle orbit, $J$ ($T$) denotes the
total angular momentum (isospin) of two nucleons. For each system 1000 runs of
calculations are  performed   in order to accumulate stable statistics. $N_p$
and $N_n$ refer to the number of valence protons and neutrons, respectively.

This paper is organized as follows. In \secref{Parity} we present our results
for parity distributions for a variety of systems. In \secref{Senior} we
discuss the distribution of seniority in the ground states using random
interactions. In \secref{Quad} we show our results for spectroscopic quadrupole
moment in the ground states which suggests  prolate or oblate shapes. In
\secref{Alpha} we discuss the $\alpha$-clustering  correlation in the ground states. The
summary will be given in \secref{DS}.

\section{\seclab{Parity}Parity}

We select four model spaces for studying the parity distribution in the ground
states obtained by random interactions:
\begin{itemize}
\item[A] Both protons and neutrons are in the
$f_{\frac{5}{2}} p_{\frac{1}{2}} g_{\frac{9}{2}}$ shell which corresponds to
nuclei with both proton number Z and neutron number N $\sim 40$;
\item[B] Protons in the
$f_{\frac{5}{2}} p_{\frac{1}{2}} g_{\frac{9}{2}}$ shell and neutrons in the
$g_{\frac{7}{2}} d_{\frac{5}{2}}$ shell which correspond to nuclei with
Z$\sim40$ and $N\sim50$;
\item[C] Both protons and neutrons are in the $h_{\frac{11}{2}}
s_{\frac{1}{2}} d_{\frac{3}{2}}$ shell which correspond to nuclei with Z and
N$\sim82$;
\item[D] Protons in the $g_{\frac{7}{2}} d_{\frac{5}{2}}$ shell and
neutrons in the $h_{\frac{11}{2}} s_{\frac{1}{2}} d_{\frac{3}{2}}$ shell which
correspond to nuclei with Z$\sim50$ and $N\sim82$.
\end{itemize}
These four model spaces do
not correspond to a complete major shell but have been truncated in order to
make the calculations feasible. These truncations are based on the sub-shell
structures for the involved single-particle levels. We study the dependence on
valence-proton number $N_p$ and valence-neutron number $N_n$ in these four
model spaces. It is noted that the number of states (denoted as $D(I)$)
 for positive and negative parity are very close
for all these examples. The  $D(I)$'s for a few examples are shown in
\figref{1}. One thus expects that  the probability of  the ground states with
positive parity is around 50$\%$, if one assumes that  each state of the full
shell model space is
 equally probable in the ground state.

\begin{figure}[tbp]
\includegraphics[bb=60 75 515 757,angle=-90,width=8cm,clip]{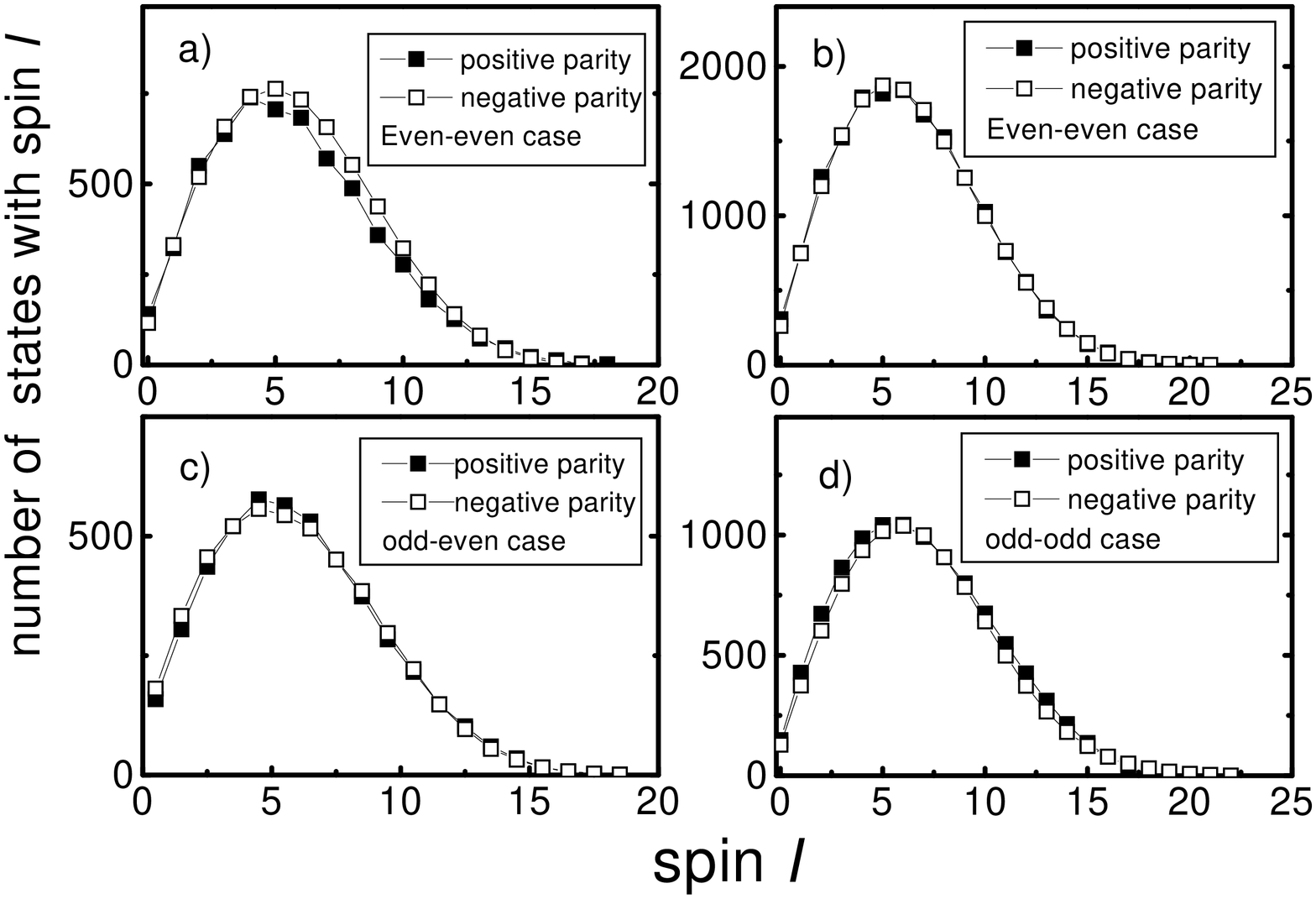}
\caption[fig1]{ Number of states with total angular momentum  $I$ (denoted as $D(I)$)
versus  $I$.
 One sees that  the $D(I)$ of positive parity levels
 and that of negative parity levels are very close to each other.
(a) ~ two protons in the $1g_{9/2} 2p_{1/2} 1f_{5/2}$ shell and four neutrons
in the $2d_{5/2} 1g_{7/2}$ shell; ~~ (b) ~ two protons and two neutrons in the
$1g_{9/2} 2p_{1/2} 1f_{5/2}$ shell; ~~ (c) ~ two protons and three neutrons in
the $1g_{9/2} 2p_{1/2} 1f_{5/2}$ shell; ~~ (d) ~ three protons in the
$1h_{11/2} 3s_{1/2} 2d_{5/2}$ shell and three neutrons in the $2d_{5/2}
1g_{7/2}$ shell. }
\figlab{1}
\end{figure}

\begin{table}
\caption{\tablab{par} The positive parity probability
for the ground states (in \%). In brackets the
number of neutrons and protons $(N_p, N_n)$ is given for each configuration.}
\begin{ruledtabular}
\begin{tabular}{cccccccc}
\multicolumn{2}{c}{basis A}\\
$(0,4)$& $(0,6)$& $(2,2)$& $(2,4)$& $(2, 6)$\\
 $86.8\%$& $86.2\%$& $93.1\%$&  $81.8\%$& $88.8\%$\\ \hline
 $(2,3)$&  $(1,4)$& $(1,3)$& $(0,5)$& $(1,5)$& $(6,1)$& $(2, 1)$\\
 $42.8\%$& $38.6\%$& $77.1\%$& $45.0\%$&    $69.8\%$& 38.4$\%$& $31.2\%$ \\\hline
\multicolumn{2}{c}{basis B}\\
$(2,2)$& $(2,4)$& $(4,2)$  \\
 $72.7\%$& $80.5\%$& $81.0\%$ \\\hline
 $(3,4)$& $(3,3)$& $(2,3)$&$(5,1)$& $(3,2$)& $(4,1)$& $(1,4)$& $(5,0)$\\
 $42.5\%$& $74.9\%$&$72.4\%$& $42.9\%$& $39.1\%$& $75.1\%$&  $26.4\%$ & $44.1\%$  \\\hline
\multicolumn{2}{c}{basis C}\\
 $(2,2)$& $(2,4)$& $(4,0)$& $(6,0)$ \\
 $92.2\%$& $81.1\%$& $80.9\%$&  $82.4\%$ \\\hline
 $(1,3)$& $(1,5)$&$(2,3)$& $(5,0)$&  $(4,1)$\\
 $73.0\%$& $64.4\%$& $52.0\%$& $42.6\%$& $56.5\%$ \\\hline
\multicolumn{2}{c}{basis D}\\
$(2,2)$& $(4,2)$& $(2,4)$& $(0,6)$ \\
 $67.2\%$& $76.1\%$& $74.6\%$&  $83.0\%$  \\\hline
 $(3,3)$& $(3,2)$& $(2,3)$& $(0,5)$\\
 $54.5\%$& $54.2\%$&$54.0\%$&  $45.9\%$\\
\end{tabular}
\end{ruledtabular}
\end{table}

The calculated statistics for the parity of the ground states, using a TBRE
Hamiltonian, is given in \tabref{par}. This clearly shows that positive parity
is favored, and dominant for most examples, for the ground states of even-even
nuclei in the  presence of  random interactions.

The statistics for nuclei  with  odd mass numbers and those with odd values for
both $N_p$ and $N_n$ is also given in \tabref{par}. This statistics shows  that
the probabilities to have positive or negative parity in the ground states are
almost equal with some exceptions. In general, there is no favoring for either
positive parity or negative parity in the ground states of odd mass nuclei and
doubly odd nuclei in the presence of random interactions. It is noted that
these calculations are done for the beginning of the shell. For the end of the
shell the results show a similar trend.

We also find that the above regularities for parity distributions also hold for
very simple cases: single-closed two-$j$ shells with one positive-parity and
one with negative parity. We have checked explicitly the cases for $(2j_1,
2j_2)$ = $(9,7)$, $(11,9)$, $(13,9)$, $(11,3)$, $(13,5)$, $(19,15)$, $(7,5)$,
$(15,1)$. The statistics is very similar to the above results: The probability
of ground states with positive parity is about $85\%$ for an even number of
nucleons and about $50\%$ for an odd number of nucleons.

It is interesting to note that for all even-even nuclei the $P(0^+)$ is usually
two orders in magnitude larger than $P(0^-)$. It would be very interesting to
investigate the origin of the large difference for $P(0)$ for positive and
negative parity states, i.e., why the $0^-$ is not favored in the ground
states. As is the case for an odd number of bosons  with spin $l$~\cite{Zhaox},
spin $I=0$ is {\it not} a sufficient condition to be favored
in the ground states of a
many-body system in the presence of the random interactions.
It should be noted that for the quantum numbers of a realistic
g.s. not only $I=0$ is required but also positive parity.

One simple and schematic system to study the parity distribution of the ground
states in the presence of random interactions is the $sp$-bosons system. First,
we note that a $sp$-boson system with an odd value of  $n$ has the same number
of states with positive and negative parity; while for an even value of $n$
there are {\it slightly} more states with positive parity (the difference is
only $n+1$, $n$ is the number of $sp$ bosons of the system). The calculated
results of Ref.~\cite{Bijker6} showed that when the number $n$ of $sp$ bosons
is even, the dominant $I$ in the ground states is 0 or $n$ (about $99\%$), with
positive parity (parity for $sp$ bosons is given by $(-)^I$) dominance. When
the number of $n$ is odd, only about 50$\%$ of the ground states in the
ensemble have spin-0, and  about $50\%$ have $I=1$ or $I=n$ ground states. This
leads to about equal percentages for positive and negative parity ground
states. This pattern is very similar to that observed for fermion systems.

\section{\seclab{Senior}seniority}

In this section, we discuss the distribution of the seniority, the number of
particles not pairwise coupled to angular momentum 0, of the ground states of
nuclei in the $sd$ shell in the presence of random interactions. Because
seniority is used in classifying the states in our basis, we define the
expectation value for seniority in the ground states as follows \cite{Auerbach}
\begin{equation}
\langle v \rangle = \sum_i f^2_i v_i ~,
\end{equation}
where $f_i$ is the amplitude of the $i-$th component in the ground state wave
function, and $v_i$ is the seniority number of the corresponding component.

For even-even nuclei we consider the spin-0 g.s.\ because  previous
discussions~\cite{Zhao-3,Johnson2} were focused on spin-$0$ ground states. For
odd-mass nuclei we consider  the $I=$1/2, 3/2, and 5/2 ground states, because
these spin $I$'s are equal to the angular momenta of the single-particle levels
in the $sd$ shell and are favored as the ground states in the presence of
random interactions. For odd-odd nuclei in this section we consider the ground
states with  $I=1$ (most favored) and $I=0$ states. The examples that we have
calculated include $(N_p, N_n)$=$(0,4)$,  $(0,6)$,  $(2,2)$,  $(2,4)$, $(2,6)$,
$(4,6)$, $(0,5)$, $(2,3)$,  $(2,5)$, $(4,3)$, $(4,5)$,  $(3,3)$, $(1,5)$,
$(3,5)$.

\begin{figure}[tbp]
\includegraphics[bb=76 33 545 753,angle=-90,width=8cm,clip]{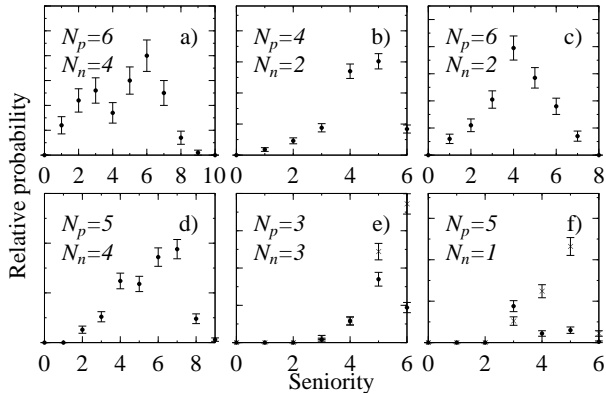}
\caption[fig2]{Distribution of seniority
in the ground states with spin zero for even-even nuclei (refer
to  panels (a), (b), and (c) ),
 spin $I=j_1, ~j_2, ~j_3$ for odd-$A$ case (refer to  panel
 (d) ) or spin $I=1$, 0
 for odd-odd nuclei (refer to panels (e) and (f) ).
The error bar is  defined by the square root of
 the count (statistics) for each seniority bin
 (step width is 1).
  The dominance of  seniority zero components
 of ground states is not observed.}
\figlab{2}
\end{figure}

Typical examples of the distribution of average seniority ($\langle v \rangle$)
in the ground states are shown in \figref{2} in arbitrary units (i.e.\ relative
probability). The figure shows that for none of the cases a small value for
$\langle v
\rangle$ is preferred. These distributions of seniority in the ground states
show that the large amplitude of for $S$-pair operator between the spin-0 g.s.\
of a $n$-nucleon system and that of a $(n+2)$-nucleon system, as observed in
Ref.~\cite{Johnson2}, should not be understood as an indication of large $S$
pair condensate in the spin-0 g.s.\ of TBRE Hamiltonians. Further studies are
necessary to understand the implications of Ref.~\cite{Johnson2}.

\section{\seclab{Quad}Spectroscopic quadrupole moment}

In this section we study the  quadrupole moments $Q$ of the ground
or low-lying states. If the ground state spin $I$ is smaller than one (i.e., 0
or 1/2), the external quadrupole-moment necessarily vanishes (even though there
could be a finite intrinsic moment) because the triangle relationship of
angular momentum coupling cannot be obeyed by the two $I$'s ($I \le
\frac{1}{2}$)
and the angular momentum for quadrupole operator (=2).
For these cases one can use an alternative, namely,  the
 quadrupole moment of the next lowest state with $I>\frac{1}{2}$.
 For all cases that we have checked, it is found that  the
essential statistics for positive
and negative  quadrupole moments
obtained by this alternative
 is  very close to those obtained by
 neglecting cases with ground state $I<1/2$.
In this paper we  show the statistics which do not includes
the runs of spin$-$0 and spin$-$1/2 ground states.  The total number
of calculated  external  quadrupole moments
is thus much less than 1000. We note
that a negative external quadrupole moment implies
a positive quadrupole moment
in the intrinsic frame  and thereby a prolate deformation.

The external quadrupole moment is   defined by
\begin{equation}
Q = \langle \beta I |r^2 Y_{2M} | \beta I \rangle
  \eqlab{quadrupole}
\end{equation}
for both proton and neutron degrees of freedom. In \eqref{quadrupole} $|
\beta I \rangle $ is the wave function of the ground state. In this paper ``Q"
will be used to refer to the external quadrupole moment following from
\eqref{quadrupole}.

\begin{table}
\caption{\tablab{3} The number of cases with positive (negative) external
quadrupole moments are given in {\bf bold} ({\it italic}) font, respectively.
We omitted the cases for which the spin of the ground state is less than 1, see
the text for further details.}
\begin{ruledtabular}
\begin{tabular}{c|cc|cc|cc}
  \multicolumn{7}{c}{both protons and neutrons in the $sd$ shell}     \\  \hline
\multicolumn{1}{c|}{ $(N_p, N_n)$} & \multicolumn{2}{c|}{(2, 1)} & \multicolumn{2}{c|}{(2, 3)} &
 \multicolumn{2}{c}{ (2,5)}  \\
 & {\bf 280} & {\it 418}
 & {\bf 338} & {\it 430}  & {\bf 306} & {\it 402} \\
\multicolumn{1}{c|}{ $(N_p, N_n)$} & \multicolumn{2}{c|}{(2, 1)} & \multicolumn{2}{c|}{(4, 3)} &
 \multicolumn{2}{c}{ (4,5)}  \\
 & {\bf 287} & {\it 425} & {\bf 434} & {\it 374}  & {\bf 320} & {\it 370} \\
\multicolumn{1}{c|}{ $(N_p, N_n)$} & \multicolumn{2}{c|}{(6, 1)} & \multicolumn{2}{c|}{(6, 3)} &
 \multicolumn{2}{c}{ (6,5)}  \\
 & {\bf 201} & {\it 530} & {\bf 400} & {\it 444}  & {\bf 420} & {\it 348} \\ \hline
  \multicolumn{7}{c}{Basis (A);  protons and neutrons in  $f_{\frac{5}{2}} p_{\frac{1}{2}}
g_{\frac{9}{2}}$     }     \\  \hline
\multicolumn{1}{c|}{ $(N_p, N_n)$} & \multicolumn{2}{c|}{(1, 2)} & \multicolumn{2}{c|}{(1, 3)} &
 \multicolumn{2}{c}{ (1,4)}  \\
 & {\bf 267} & {\it 469}
 & {\bf 283} & {\it 481}
 & {\bf 246} & {\it 535} \\
\multicolumn{1}{c|}{ $(N_p, N_n)$} & \multicolumn{2}{c|}{(1, 6)} & \multicolumn{2}{c|}{(2, 3)} &
 \multicolumn{2}{c}{ (0,5)}  \\
 & {\bf 207} & {\it 566} & {\bf 284} & {\it 564} & {\bf 447} & {\it 459} \\ \hline
\multicolumn{7}{c}{ Basis (B); protons ($f_{\frac{5}{2}} p_{\frac{1}{2}}
 g_{\frac{9}{2}}$), neutrons  ($g_{\frac{7}{2}} d_{\frac{5}{2}}$) } \\ \hline
\multicolumn{1}{c|}{ $(N_p, N_n)$} & \multicolumn{2}{c|}{(1, 4)} & \multicolumn{2}{c|}{(4, 1)} &
 \multicolumn{2}{c}{ (2, 4)}  \\
 & {\bf 374} & {\it 507}
 & {\bf 278} & {\it 632}
 & {\bf 253} & {\it 428} \\
\multicolumn{1}{c|}{ $(N_p, N_n)$} & \multicolumn{2}{c|}{(3, 4)} & \multicolumn{2}{c|}{(4, 3)} &
 \multicolumn{2}{c}{ (6, 1)}  \\
 & {\bf 278} & {\it 620}
 & {\bf 330} & {\it 560}
 & {\bf 233} & {\it 660} \\ \hline
  \multicolumn{7}{c}{Basis (C); protons and neutrons in  $s_{\frac{1}{2}} d_{\frac{3}{2}}
h_{\frac{11}{2}}$    }     \\  \hline
\multicolumn{1}{c|}{ $(N_p, N_n)$} & \multicolumn{2}{c|}{(2, 3)} & \multicolumn{2}{c|}{(2, 5)} &
 \multicolumn{2}{c}{ (4,3)}  \\
 & {\bf 231} & {\it 657} & {\bf 238} & {\it 472}  & {\bf 392} & {\it 498} \\
\multicolumn{1}{c|}{ $(N_p, N_n)$} & \multicolumn{2}{c|}{(5, 1)} & \multicolumn{2}{c|}{(5, 0)} &
 \multicolumn{2}{c}{ (3,3)}  \\
 & {\bf 213} & {\it 628}  & {\bf 212} & {\it 659}  & {\bf 349} & {\it 449} \\\hline
  \multicolumn{7}{c}{Basis (D); protons $g_{\frac{7}{2}} d_{\frac{5}{2}}$,
neutrons   $s_{\frac{1}{2}} d_{\frac{3}{2}} h_{\frac{11}{2}}$  }     \\  \hline
\multicolumn{1}{c|}{ $(N_p, N_n)$} & \multicolumn{2}{c|}{(14, 13)} & \multicolumn{2}{c|}{(15, 12)} &
 & \\
 & {\bf 781} & {\it 183} & {\bf 610} & {\it 333} & &\\
\end{tabular}
\end{ruledtabular}
\end{table}

We have calculated ``Q" for a number of cases in the  $sd$ shell and for
several  fillings of the four single-particle bases mentioned in
\secref{Parity}. The results are given in
\tabref{3}. One sees that negative values for $Q$ (corresponding
to prolate deformations) are dominant with two exceptions, ($N_p, N_n$)=(4,3),
(6,5) in the $sd$ shell. In general we observe that for the $sd$ shell the
statistics for positive and  negative values for $Q$ are comparable if $N_p$
and/or $N_n$ are close to their mid-shell values.

From \tabref{3} we  conclude  that  at the beginning of the shell negative
values for $Q$ are dominant,  while at the end of the shell  positive  value
dominate. This is  similar to the result  for a harmonic oscillator potential,
for which prolate deformation occurs at the beginning and oblate deformation at
the end of the shell~\cite{Mottelson}.

\section{\seclab{Alpha} $\alpha$-clustering}

It was shown in Ref.~\cite{Arima2} that the essential parts of the $I=0$, $T=0$
ground state for $^{20}$Ne with two protons and two neutrons in the $sd$ shell
is dominated by components with highest orbital symmetry, $[4]$; $91.8\%$ of
the ground state is given by components with orbital symmetry  $\left[4
\right]$ which corresponds to a pure $\alpha$-clustering configuration.
One may use the expectation value of the Majorana interaction, $P_M$, as the
fingerprint for the $\alpha$-cluster wave function.  Another similar example is
the $I=0$, $T=0$ ground state  for $^{8}$Be with two protons and two neutrons
in the $p$-shell.  If one uses Cohen-Kurath interaction, one sees that the
expectation value of $P_M$ is $-5.76$, close to $-6$ which is the eigenvalue of
Majorana force.  The overlap between the g.s.  wave function obtained by
diagonalizing the Cohen-Kurath interaction for $^{8}$Be and that for
exact SU(4) symmetry (namely, full  symmetry $\left[ 4 \right]$ for the ground
state) is $0.97$. This dominance full symmetry $[4]$ in the $I=0$ and $T=0$
ground state of these nuclei with respect to the permutation of orbital degree
of freedom is an indication of $\alpha$-clustering  correlation  from the
perspective of the shell model. In this paper  we concentrate on these two
examples using random interactions.

To set the scale, we can  calculate the matrix element of $P_M$ in the
$I=T=0$ (spin-isospin singlet) ground state by assuming that all the possible
$I=T=0$ states with different symmetries with respect to exchanging the orbital
degree of freedom appear with equal-probably. We call the $P_M$ such obtained
the geometric $P_M$. To do so one needs the number of ($I=0$, $T=0$) states for
each symmetry of orbital degree.

The procedure to construct  the states with particular spin-isospin symmetry is
given in Ref.~\cite{Hamermesh}, while that for constructing wave functions with
certain orbital symmetry is given in Ref. \cite{Harvey} for the study the
Elliott model \cite{Elliott}, with tables for the $sd$, $pf$ and $sdg$ shells.
Finally, the spin-isospin functions should be coupled to the orbital functions
with their conjugate symmetry to obtain the fully anti-symmetric wave functions
with respect to exchange two particles. The angular momentum for each state is
given by coupling $S$ and $L$.

\tabref{2}
presents the number of  $I=0$ states for two protons
and two neutrons in the $p$ shell and the $sd$ shell with
all possible symmetries with respect to orbital degree of freedom.
From Table II one  obtains the geometric  $P_M$ for the $I=T=0$ states:
 the geometric $P_M$ is $-\frac{6}{5}$ for the $p$ shell and
$-\frac{22}{21}$ for the $sd$ shell.

\begin{table}
\caption{\tablab{2}
The number of $I=0$ states for two valence  protons and two
valence neutrons in the $p$ shell and the $sd$ shell with definite
symmetry with respect to exchange orbital degree of freedom of
two particles and the corresponding conjugate symmetry with respect
to exchange  spin-isospin ($S-T$) degrees of freedom.
$L$ is the total orbital angular momentum,
and $S$ is the total spin. The last column gives  number of the $I=0$
states with  $T=0$.}
\begin{ruledtabular}
\begin{tabular}{llcc}
$L$  &  $S$   &  $I=0$  &  $ I=T=0$  \\  \hline
  \multicolumn{4}{c}{the $p$ shell}     \\  \hline
$\left[4 \right]$ ~~ $ 0, 2, 4$ &    $ 0$  &  $0$ & $0$    \\
$\left[31 \right]$ ~ $ 1, 2, 3$ &    $0, 1^2$  &   $0^2$ & $0$  \\
$\left[22 \right]$ ~ $ 0, 2$ &    $ 0^2, 1, 2$  &   $0^3$ & $0^2$  \\
$\left[211 \right]$ $ 1$ &    $ 0, 1^3, 2$  &   $0^3$ & $0$  \\  \hline
  \multicolumn{4}{c}{the $sd$ shell}     \\  \hline
$\left[4 \right]$ ~~~ $  0^4, 2^5, 3, 4^4, 5, 6^2, 8 $ &     $ 0$  &  $0^4$ & $0^4$    \\
$\left[31 \right]$ ~ $  0^2, 1^4, 2^7, 3^6, 4^5, 5^3, 6^2, 7$ &     $ 0, 1^2$  &   $0^{10}$ & $0^4$  \\
$\left[22 \right]$ ~ $  0^3, 1, 2^5, 3^2, 4^3, 5, 6  $ &    $  0^2, 1, 2$  &   $0^{12}$ & $0^8$  \\
$\left[211 \right]$  $ 1^5, 2^3, 3^5, 4^2, 5^2$ &     $  0, 1^3, 2$  &   $0^{18}$ & $0^{5}$  \\
$\left[1111 \right]$ $ 1, 2, 3$ &     $ 0, 1, 2$  &   $0^2$ &  -
\end{tabular}
\end{ruledtabular}
\end{table}

Using a TBRE Hamiltonian we obtain the following probabilities for  spin--$I$
ground states: For 1000 runs, one obtains 485 and 365 runs
with $(I,T)=(0,0)$ ground states for $^{8}$Be and $^{20}$Ne, respectively.
This is consistent with  the result~\cite{Johnson1,Johnson2} of the $I=T=0$
g.s.\ dominance in the presence of random interactions. The average value of
$P_M$ for the  $(I,T)=(0,0)$ g.s.\ that we obtain is $-1.26$ (the geometric
value is $-\frac{6}{5}=-1.20$) and $-1.66$ (the geometric value is
$-\frac{22}{21}=-1.05$) for the $p$ shell and the $sd$ shell, respectively. The
average value of $P_M$ for a TBRE Hamiltonian and the corresponding geometric
value are very close for the $p$ shell, indicating that $\alpha$-clustering
correlation is not favored by random interactions. For the case of  the $sd$ shell, the
average value of $P_M$ for a TBRE Hamiltonian deviates sizably from  its
geometric value.

To check whether this deviation becomes larger for larger shells, we calculate
the case of two protons and two neutrons in the $sdg$ shell, for which we
obtained 385 cases with $(I,T)=(0,0)$ ground states among 1000 sets of the TBRE
Hamiltonians.
 The average $P_M$ values
for these states is $-0.629$ while that by assuming a random orbital symmetry
is $-\frac{2}{5}$, which is close.

It is also interesting to study the distribution of overlaps between the
$I=T=0$ ground state obtained from the realistic interactions and those obtained
by pure random interactions or by a combination of realistic  and
random interactions. As an example we discuss here the case of two protons and
two neutrons in the $p$-shell where the realistic interaction is chosen as the
Cohen-Kurath interaction. We thus define a Hamiltonian
\begin{equation}
H = (1-\lambda) H_{\rm TBRE} + \lambda H_{\rm real} ~.
\eqlab{real}
\end{equation}
Here $\lambda =0$ corresponds to the pure TBRE Hamiltonian and $\lambda =1$
corresponds to the realistic Cohen-Kurath interaction. We will vary $\lambda $
in the range from 0 to 1, corresponding to the situation of nuclear forces with
different  mixtures of random noise.

\begin{figure}[tbp]
\includegraphics[width=8cm,bb=15 15 500 730,clip]{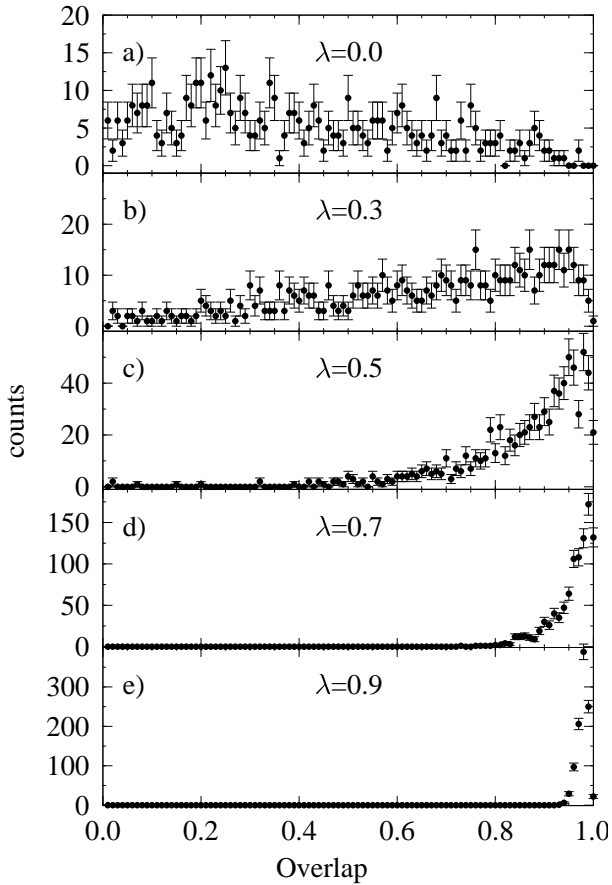}
\caption[fig3]{ The overlaps between the $I=T=0$ ground states for two protons and
two neutrons in the $p$-shell  obtained by Cohen-Kurath interactions  and those
obtained by the Hamiltonian \eqref{real}. (a)-(e) corresponds to $\lambda =0,
0.3, 0.5, 0.7, 0.9$, respectively. }
\figlab{3}
\end{figure}

The results for $\lambda =0, 0.3, 0.5, 0.7$ and 0.9 are shown in \figref{3}
(a)-(e). The error bars indicate the statistical error in determining the
numbers, defined by the square root of
 the number of counts for each bin.
For case (a) with $\lambda =0$ one sees that the overlaps distributes
``randomly" from 0 to 1.  This suggests that  pure random interactions produce
``random" overlaps with respect to the realistic interactions. However, for
$\lambda>0.5$, the $I=T=0$ ground states are close to that of the realistic
interactions for most of the cases. This is especially clear from \figref{4}
where the overlap, averaged over the different Hamiltonians in the ensemble
\eqref{real} is plotted versus $\lambda$. The statistical inaccuracies are
indicated by the error bars in this figure. For values of $\lambda$ exceeding
0.6 the overlap is very close to unity while for larger admixtures of the
random component in the interaction the overlap decreases approximately
linearly with $\lambda$. This trend has all the signatures of a second-order
phase transition. Only for limited magnitude of the random interaction the
g.s.\ has a realistic structure which breaks down when a critical value is
exceeded.

\begin{figure}[tbp]
\includegraphics[width=6cm,bb=10 130 355 435,clip]{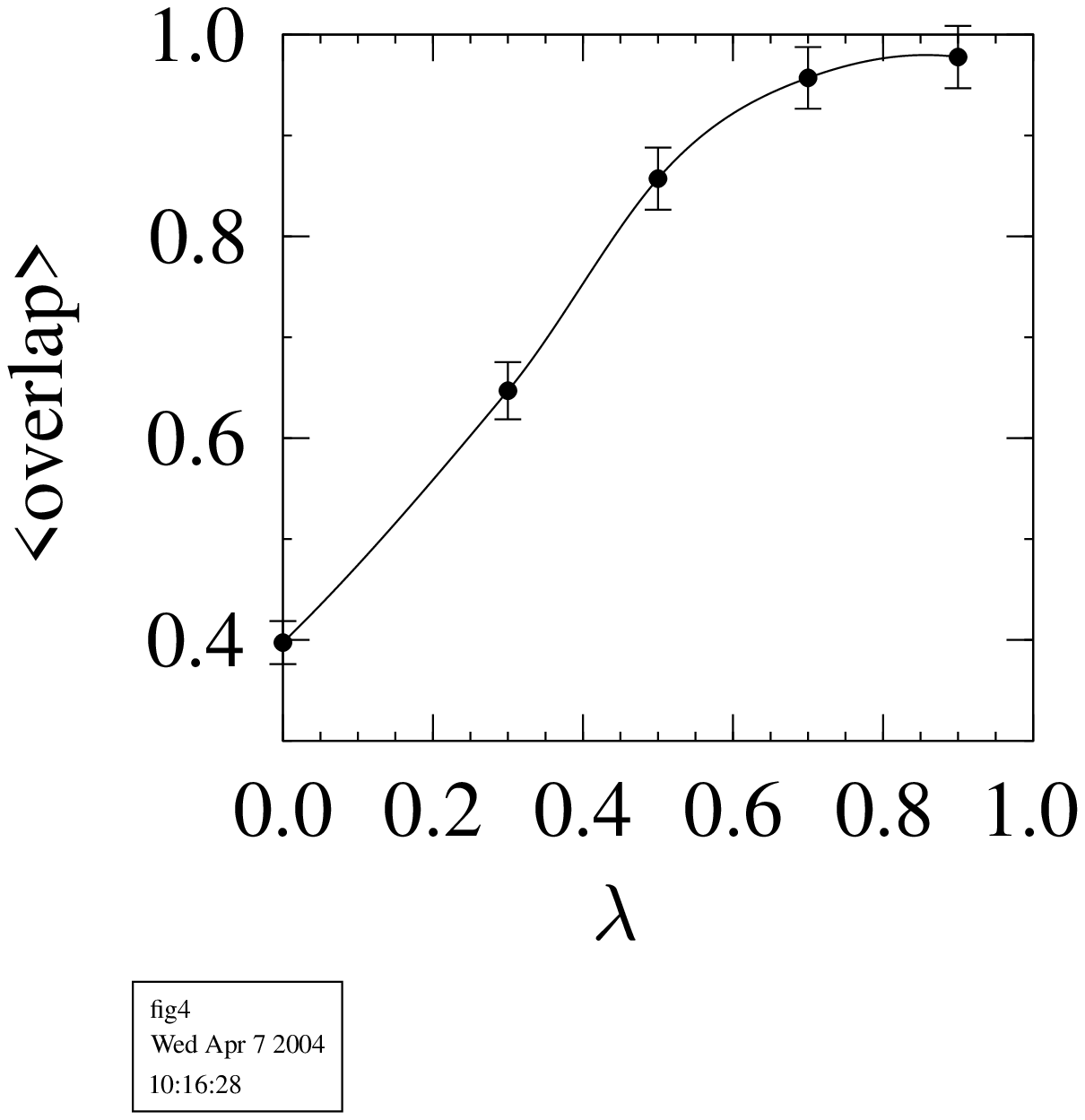}
\caption[Ratio]{Average overlap of the g.s.\ of the Hamiltonian of \eqref{real}
with that of the realistic Hamiltonian as function of the mixing parameter
$\lambda$. The line is plotted to guide the eyes. }
\figlab{4}
\end{figure}

\section{\seclab{DS}Discussion and Summary}

The present paper was stimulated by the discovery of the spin 
zero ground state dominance (0 g.s.) of even fermion systems \cite{Johnson1}
and boson systems  \cite{Bijker0}  in the presence of  the random 
two-body ensemble (TBRE).  This discovery sparked off a sudden 
interest of many-body systems under the TBRE.  It also led to extensive studies 
such as energy centroid of fixed spin states, collectivity, etc.    The 
purpose of this paper is to study the robustness of some features which 
are well known in nuclear physics but have not been studied under 
the TBRE. 

First, we calculated in Sec. II the parity distribution of 
the g.s.\ for a TBRE 
Hamiltonian. It was found that positive parity is dominant for the g.s.\ of
systems with even numbers of valence protons and neutrons. For odd-$A$ and
doubly odd systems, the TBRE Hamiltonian leads to ground states with comparable
probability for both positive  and negative parity.  This is similar to
the global statistics for the realistic nuclei
with $A\ge 70$ (refer to Table I). Unlike  the spin-0 g.s.\  dominance in
the presence of random interactions, the dominance of positive parity in the
ground states of even-even nuclei has not been pointed out explicitly. Since  parity is a
much simpler quantity than angular momentum, an understanding of the parity
dominance of even-even systems may be helpful  in understanding the spin-0
g.s.\ dominance  of even-even nuclei in the presence of random interactions.

Second,  our investigation showed that the seniority distribution for
the g.s.\ of $sd$-shell nuclei is not dominated by low seniority components,
contrary to the situation for realistic nuclei. Our investigation also suggests
that the correlation  between the wave function of the spin-0 g.s.\ for $A$
nucleons and that for $A+2$ nucleons discussed  in Ref.~\cite{Johnson2} should
not be understood as an indication of seniority zero component dominance.

Third, the dominance of negative external quadrupole moments at the beginning of the shell and
positive quadrupole moments at the end of the shell is also observed in the
g.s.\ obtained by using the TBRE interactions. This situation is similar to the
prediction obtained from a simple harmonic oscillator potential. This means
that the TBRE Hamiltonians does not lead to an overall dominance of the prolate
deformation, however also in nuclei a dominance of prolate deformation is
observed when both valence protons and neutrons are in the first half of a
major shell.

Last, we studied the $\alpha$-clustering correlation by calculating the
expectation value of the Majorana operator in the $I=0$, $T=0$ g.s.\ which are
obtained using TBRE interactions.  We also calculated the overlaps 
between the $I=0$, $T=0$ ground states obtained by using 
the TBRE Hamiltonian and the ground state by using realistic interactions. 
Our calculations on $^{8}$Be and $^{20}$Ne
showed that the $\alpha$-clustering structure is not favored by a pure TBRE
Hamiltonian. It is interesting to note that as function of the admixture of a
realistic Hamiltonian a second-order phase transition is observed. For
Hamiltonians that contain less than $\sim 0.4$ admixture of  random interactions,   the
structure of the g.s.\ is close to realistic, but for higher admixtures the overlap with
a realistic wave functions becomes progressively worse.

In conclusion,  we have observed in this paper
the  dominance of positive parity in the ground states
of even-even nuclei in the presence of pure random two-body interactions.
Because parity is an intrinsically simpler quantum number, it will be  interesting
to understand the mechanism for this. In addition, it has been shown that,
even though the quantum numbers for the g.s.\ are realistic, the dynamic
properties of the ground states under the TBRE Hamiltonian, 
such as seniority which is a signature of pairing 
correlation,  the $\alpha$-clustering probabilities, and the sign of 
quadrupole moments, are in sharp
disagreement with those for realistic nuclei.

\begin{acknowledgments}
Part of this work was performed as part of the research program of the
Stichting voor Fundamenteel Onderzoek der Materie (FOM)
with financial support
from the Nederlandse Organisatie voor Wetenschappelijk
Onderzoek (NWO).
This work was also supported in part by
 a Grant-in-Aid for Specially Promoted  Research
(Grant No. 13002001) from the Ministry of
Education, Science  and Culture in Japan.  One of
the authors (Y.M.Z.) acknowledges a grant from the NWO.

\end{acknowledgments}

\end{document}